\documentstyle[12pt]{article}

\def\+{{+\!\!\!+}}
\def\pp{\mbox{\tiny${}_{\stackrel\+ =}$}}
\def\mm{\mbox{\tiny${}_{\stackrel= \+}$}}

\def\d{\partial}

\def\bR{\hbox{I\hspace{-0.02in}R}} 
\def\pmb#1{\setbox0=\hbox{#1}%
  \kern.0em\copy0\kern-\wd0
  \kern-.04em\copy0\kern-\wd0
  \kern.08em\copy0\kern-\wd0
  \kern-.04em\raise.0433em\box0 }       

\newcommand{\nc}{\newcommand}
\nc{\beq}{\begin{equation}}
  \nc{\eeq}[1]{\label{#1}\end{equation}}
\nc{\ber}{\begin{eqnarray}}
  \nc{\eer}[1]{\label{#1}\end{eqnarray}}
\nc{\pek}[1]{\cite{#1}}
\nc{\enr}[1]{(\ref{#1})}
\nc{\kal}[1]{{\cal{#1}}}
\nc{\dott}{\;\cdot\;}

\newcommand{\be}{\begin{equation}}
  \newcommand{\ee}{\end{equation}}
\newcommand{\bea}{\begin{eqnarray}}
  \newcommand{\eea}{\end{eqnarray}}

\newcommand{\Section}[1]{\section{#1} \setcounter{equation}{0}}

\begin{document}

\begin{center}

  \begin{flushright}
    UUITP-05-03\\
    USITP-2003-02\\
    hep-th/0304013\\
  \end{flushright}

  \vskip .3in \noindent

  \vskip .1in

  {\large \bf{Superconformal boundary conditions for the WZW model}}
  \vskip .2in

  {\bf Cecilia Albertsson}$^a$\footnote{e-mail address: cecilia@physto.se},
  {\bf Ulf Lindstr\"om}$^b$\footnote{e-mail address: ulf.lindstrom@teorfys.uu.se}
  and  {\bf Maxim Zabzine}$^{c}$\footnote{e-mail address: zabzine@fi.infn.it} \\

  \vskip .15in

  \vskip .15in
  $^a${\em  Institute of Theoretical Physics, \\
    AlbaNova University Center, \\
    University of Stockholm,
    SE-106 91 Stockholm, Sweden}\\
  \vskip .15in
  $^b${\em  Department of Theoretical Physics, \\
    Uppsala University,
    Box 803, SE-751 08 Uppsala, Sweden}\\
  \vskip .15in
  $^c${\em INFN Sezione di Firenze, Dipartimento di Fisica, \\
    Via Sansone 1, I-50019 Sesto F.no (FI), Italy}

  \bigskip

  \vskip .1in

\end{center}
\vskip .4in

\begin{center} {\bf ABSTRACT }
\end{center}
\begin{quotation}\noindent
  We review the most general, local, superconformal boundary
  conditions for the two-dimensional $N=1$ and $N=2$ non-linear sigma
  models, and analyse them for the $N=1$ and $N=2$ supersymmetric WZW
  models. We find that the gluing map between the left and right
  affine currents is generalised in a very specific way as compared to
  the constant Lie algebra automorphisms that are known.
\end{quotation}
\vfill
\eject

\Section{Introduction}

In recent years D-branes on group manifolds have attracted a great
deal of attention as an ideal setting for the study of D-branes in
general string backgrounds.  Because the world-sheet theories
corresponding to strings on a group manifold are solvable
Wess-Zumino-Witten (WZW) conformal field theories, one can analyse the
stringy effects in great detail.  WZW models have an underlying affine
symmetry which corresponds to affine Lie algebras.  By now D-branes
associated with an automorphism of the affine algebra are fairly well
understood, at least in the compact case
\cite{Alekseev,Schomerus:2002dc}.  Different properties of these
branes such as stability \cite{Bachas}, underlying geometry
\cite{Felder,Stanciu:1999id}, effective actions
\cite{Alekseev5,Alekseev4,Alekseev3,Alekseev2} etc., have been
extensively studied.

Despite this progress there is a wide class of D-branes (i.e., boundary
conditions) of WZW models which are less well understood and which
correspond to the preservation of only (super)conformal symmetry. The
only models which are rational with respect to (super)conformal
symmetry are the (super) minimal models. The compact WZW models are
rational only with respect to the affine symmetry but not with respect
to (super)conformal symmetry. Therefore it is not so straightforward
to analyse the (super)conformal branes. Interesting steps in this
direction have been taken in \cite{Gaberdiel:2001xm}.

In the present letter we would like to attack the problem from a
different side.  The WZW model provides a typical example of a string
background admitting a sigma model description.  This allows one to
undertake a complementary study of the possible D-brane
configurations.  The sigma model description is valuable since it
provides a geometric interpretation of D-branes in the semiclassical
limit.  We would like to analyse classical superconformal conditions
that are local in terms of the sigma model fields and that do have
such an interpretation.  Furthermore we want to rewrite and interpret
these boundary conditions in terms of gluing conditions for the affine
currents. At the classical level, conformal symmetry is too weak a
requirement on its own to draw any conclusions. Besides, conformal
symmetry by itself does not guarantee a consistent geometrical
description. As has been shown in
\cite{Albertsson:2001dv,Albertsson:2002qc}, one must impose minimal
supersymmetry together with conformality to obtain boundary conditions
with a consistent geometrical description in terms of submanifolds.

The paper is organised as follows. We begin by reviewing the most
general local superconformal boundary conditions of the non-linear
sigma model.  In section~\ref{N1review} we review the results obtained
in \cite{Albertsson:2001dv,Albertsson:2002qc} for $N=1$, and in
section~\ref{N2review} we review the $N=2$ conditions derived in
\cite{LZ1,LZ2}.  These conditions are then analysed for the special
cases of $N=1$ and $N=2$ WZW models, in sections~\ref{N1WZW} and
\ref{N2WZW}, respectively. Finally, in section~\ref{discussion}, we
summarise and discuss the conclusions.

\Section{$\mathbf N=1$ review}
\label{N1review}

Here we give a summary of the derivation of boundary
conditions for the $N=1$ supersymmetric non-linear sigma model; for
details, see \cite{Albertsson:2001dv,Albertsson:2002qc}.
This model is given by the following action,
\begin{equation}
  \label{sigmamodel}
  S= \int\limits_{\Sigma} d^2\xi\,\,
  \left [ \d_\+ X^\mu \d_= X^\nu E_{\mu\nu}  +
    i \psi^\mu_+\nabla^{(+)}_- \psi_+^\nu
    g_{\mu\nu} + i \psi_-^\mu \nabla^{(-)}_+ \psi^\nu_- g_{\mu\nu} +
    \frac{1}{2} \psi^\lambda_+ \psi_+^\sigma \psi_-^\rho \psi_-^\gamma
    {\cal R}^{-}_{\rho\gamma\lambda\sigma} \right ],
\end{equation}
where $E_{\mu\nu}\equiv g_{\mu\nu} + B_{\mu\nu}$ denotes a background
of general (Riemannian) metric $g_{\mu\nu}$ and two-form $B$ with
field strength $H=dB$. The curvature ${\cal
  R}^{\pm}_{\rho\gamma\lambda\sigma}$ is defined as
\beq
{\cal R}^{\pm\mu}_{\,\,\,\sigma\rho\lambda} =
\Gamma^{\pm\mu}_{\,\,\,\lambda\sigma,\rho}
- \Gamma^{\pm\mu}_{\,\,\,\rho\sigma,\lambda}
+ \Gamma^{\pm\mu}_{\,\,\,\rho\gamma}
\Gamma^{\pm\gamma}_{\,\,\,\lambda\sigma}
- \Gamma^{\pm\mu}_{\,\,\,\lambda\gamma}
\Gamma^{\pm\gamma}_{\,\,\,\rho\sigma} ,
\eeq{defcur}
with $\Gamma^{\pm}$ given by
\beq
\Gamma^{\pm\nu}_{\,\,\rho\sigma} \equiv \Gamma^{\nu}_{\,\,\rho\sigma} \pm
g^{\nu\mu} H_{\mu\rho\sigma},
\eeq{definGammapm}
and $\nabla^{(\pm)}_{\pm}$ by
\beq
\nabla^{(+)}_{\pm}\psi_{+}^\nu = \partial_{\pp}\psi_{+}^\nu +
\Gamma^{+\nu}_{\,\,\rho\sigma}\d_{\pp} X^\rho
\psi_{+}^\sigma,\,\,\,\,\,\,\,\,\,\, \nabla^{(-)}_{\pm}\psi_{-}^\nu =
\partial_{\pp}\psi_{-}^\nu + \Gamma^{-\nu}_{\,\,\rho\sigma}\d_{\pp}
X^\rho \psi_{-}^\sigma .
\eeq{covdervferm}

At the classical level the model (\ref{sigmamodel}) has $N=(1,1)$
superconformal symmetry modulo boundary terms.  This symmetry
gives rise to conserved currents, namely the stress tensor $T$ and
the supersymmetry current $G^1$.  The components of the currents are
\beq
G^1_{+} = \psi_{+}^\mu \d_{\+} X^\nu g_{\mu\nu} - \frac{i}{3}
\psi_{+}^\mu \psi_{+}^\nu \psi_{+}^\rho H_{\mu\nu\rho} ,
\eeq{Gp}
\beq
G^1_{-} = \psi_{-}^\mu \d_{=} X^\nu g_{\mu\nu} + \frac{i}{3}
\psi_{-}^\mu \psi_{-}^\nu \psi_{-}^\rho H_{\mu\nu\rho} ,
\eeq{Gm}
\vspace{0.1cm}
\beq
T_{++} = g_{\mu\nu} \d_{\+}X^\mu \d_{\+}X^\nu  + ig_{\mu\nu}
\psi^\mu_+ \nabla^{(+)}_{+} \psi^\nu_{+}  ,
\eeq{Tp}
\vspace{0.02cm}
\beq
T_{--} = g_{\mu\nu} \d_{=}X^\mu \d_{=}X^\nu +
i g_{\mu\nu} \psi^\mu_- \nabla^{(-)}_{-} \psi^\nu_{-} ,
\eeq{Tm}
and they are conserved, $\d_{\pp} G^1_\mp =0$ and
$\d_{\mm} T_{\pm\pm} =0$. In the absence of a boundary,
these conservation laws give rise to four
conserved charges.
However, in the presence of boundaries one must take extra care
when deriving the conserved charges from $T_{\pm\pm}$ and
$G^1_\pm$. It turns out that the appropriate boundary conditions
to be imposed on the currents are
\begin{equation}
  \label{bulkcurbc}
  \left[ \,\, T_{++}-T_{--} \,\,
  \right]_{\sigma=0,\pi} =0,\,\,\,\,\,\,\,\,\,\,\,\,\,\,\,
  \left[ \,\,  G^1_{+}-\eta_1 G^1_{-} \,\, \right]_{\sigma=0,\pi} =0 ,
\end{equation}
where the world-sheet $\Sigma$ is assumed to be $\bR \times [0, \pi]$
(i.e., $\tau \in \bR$ and $\sigma \in [0, \pi]$), and $\eta_1 = \pm 1$
corresponds to the choice of spin structures. Our goal is to solve
conditions (\ref{bulkcurbc}) in terms of $X^\mu$ and $\psi_\pm^\mu$
such that the solution is local in these fields (the locality is
important for a geometrical interpretation of the boundary
conditions).  The most general local fermionic boundary
condition allowed by dimensional analysis has the simple form
\begin{equation}
  \label{fermans}
  \psi^\mu_- = \eta_1 R^\mu_{\,\,\nu} (X) \psi^\nu_+ ,
\end{equation}
where $R^\mu_{\,\,\nu}$(X) is a locally defined object which transforms
as a (1,1) tensor field under coordinate transformations.
The bosonic counterpart of (\ref{fermans}) can be derived
by means of a supersymmetry transformation,\footnote{For an explicit
  component form of this transformation, see appendix~A in
  \cite{Albertsson:2002qc}.} and reads
\begin{equation}
  \label{boscond}
  \partial_= X^\mu - R^\mu_{\,\,\nu}\partial_{+\!\!\!+} X^\nu
  + 2i(P^\sigma_{\,\,\rho}
  \nabla_\sigma R^\mu_{\,\,\nu} + P^\mu_{\,\,\gamma} g^{\gamma\delta}
  H_{\delta\sigma\rho} R^\sigma_{\,\,\nu})\psi_+^\rho \psi_+^\nu =0 ,
\end{equation}
where $P^\mu_{\,\,\nu} \equiv (\delta^\mu_{\,\,\nu} +
R^\mu_{\,\,\nu})/2$.

In \cite{Albertsson:2002qc} we have shown that the expressions
(\ref{fermans}) and (\ref{boscond}) are the solutions of
(\ref{bulkcurbc}) provided that
$R^\mu_{\,\,\nu}$ satisfies the following conditions,
\begin{equation}
  \label{RgRcond1}
  g_{\rho\sigma} = R^\mu_{\,\,\rho} g_{\mu\nu} R^\nu_{\,\,\sigma} ,
\end{equation}
\beq
P^\rho_{\,\,\tau} R^\mu_{\,\,\sigma} g_{\mu\nu} \nabla_\rho
R^\nu_{\,\,\gamma} + P^\rho_{\,\,\sigma} R^\mu_{\,\,\gamma} g_{\mu\nu}
\nabla_\rho R^\nu_{\,\,\tau}+ P^\rho_{\,\,\gamma} R^\mu_{\,\,\tau}
g_{\mu\nu} \nabla_\rho R^\nu_{\,\,\sigma} + 4 P^\mu_{\,\,\tau}
P^\nu_{\,\,\sigma} P^\rho_{\,\,\gamma} H_{\mu\nu\rho} = 0 .
\eeq{ALZ320}
To understand the geometrical meaning of these conditions we need to
introduce new objects.  We define a projector
$Q^\mu_{\,\,\nu}$ (i.e., $Q^2=Q$) such that $RQ=QR=-Q$. The
complementary projector $\pi=I-Q$ ($I$ is the identity operator)
satisfies $\pi P = P\pi = P$.  Then,
by contracting (\ref{ALZ320}) with $Q^\tau_{\,\,\lambda}$, we
find the condition
\begin{equation}
  \label{piinteg1}
  \pi^\mu_{\,\,\gamma} \pi^\rho_{\,\,\nu}
  Q^\delta_{\,\,[\mu , \rho]} = 0 ,
\end{equation}
which is the integrability condition for $\pi$. Therefore \emph{there is a
  maximal integral submanifold} which corresponds to the distribution
$\pi$.  We may also contract (\ref{boscond}) with $Q$, and use the
integrability condition (\ref{piinteg1}), to obtain that
$\pi^\mu_{\,\,\nu} \d_0 X^\nu = \d_0 X^\mu$ (or equivalently
$Q^\mu_{\,\,\nu}\d_0
X^\nu = 0$) and thus the end of the string is confined in the
directions transverse to a given maximal integral submanifold.

The object $R^\mu_{\,\,\nu}$ carries additional information.  We
define a $B$-field\footnote{More precisely, a gauge invariant
  combination of a $U(1)$-field strength and the background $B$-field
  $B_{\mu\nu}$.}  $B^\pi_{\mu\nu}$, i.e., a two-form living on the
submanifold corresponding to $\pi$, by
\beq
\pi_{\,\,\mu}^\rho g_{\rho\sigma} \pi^\sigma_{\,\,\nu} - B^\pi_{\mu\nu}
= ( \pi_{\,\,\mu}^\rho g_{\rho\sigma} \pi^\sigma_{\,\,\lambda} +
B^\pi_{\mu\lambda}) R^\lambda_{\,\,\nu} .
\eeq{definBpi}
Contracting (\ref{ALZ320}) with $\pi$ and using the definition
(\ref{definBpi}) we find that
\beq
(dB^\pi)_{\mu\nu\rho} = \pi_{\,\,\mu}^\lambda \pi_{\,\,\nu}^\gamma
\pi_{\,\,\rho}^\sigma H_{\lambda\gamma \sigma} .
\eeq{dBpiH}
Thus an $R^\mu_{\,\,\nu}$
which satisfies (\ref{ALZ320}) encodes both the integrable
distribution $\pi$ and the two-form $B^\pi_{\mu\nu}$ living on the
corresponding integral submanifold. In addition $dB^\pi$ coincides
with the pull-back of $H_{\mu\nu\rho}$
to this submanifold.  For details of the
analysis leading to the above results we refer the reader to
\cite{Albertsson:2001dv} and \cite{Albertsson:2002qc}.

In summary, there is a one-to-one correspondence between local
superconformal boundary conditions and submanifolds with extra
($B$-field) structure.

\Section{${\mathbf N=2}$ review}
\label{N2review}

We now turn to the $N=2$ sigma model, reviewing the
results of \cite{LZ1}.

Any Riemannian target manifold ${\cal M}$ admits an $N=(1,1)$ sigma
model. If the geometry on ${\cal M}$ is further restricted
\cite{Gates:nk} the sigma model can have $N=(2,2)$ supersymmetry.
The target manifold ${\cal M}$ of the $N=(2,2)$ sigma model
must be equipped with two complex structures $J^\mu_{\pm\nu}$,
and the metric $g$ must be Hermitian with respect to both of these.
These complex structures should be such that
\beq
l^{\mu\rho} \equiv (J^\mu_{+\nu} + J^\mu_{-\nu}) g^{\nu\rho} ,
\,\,\,\,\,\,\,\,\,\,\,\,\,\,\,\,\,
m^{\mu\rho} \equiv (J^\mu_{+\nu} - J^\mu_{-\nu})g^{\nu\rho}
\eeq{defPoisst}
are Poisson bi-vectors.  The torsion $H_{\mu\nu\rho}$ is then
defined via the Schouten bracket of these two Poisson structures
\cite{Lyakhovich:2002kc}.

In addition to the currents (\ref{Gp})--(\ref{Tm}), an $N=(2,2)$ sigma
model has two further supersymmetry currents, $G^2_{\pm}$, and two
$U(1)$ R-symmetry currents $J_{\pm}$.
In terms of world-sheet fields the currents read
\beq
G^2_{+}  = \psi_+^\mu \d_\+X^\nu J_{+\mu\nu} +
\frac{i}{3} \psi_+^\mu \psi^\nu_+
\psi^\rho_+ J^{\lambda}_{+\mu} J^\sigma_{+\nu} J^\gamma_{+\rho}
H_{\lambda\sigma\gamma} ,
\eeq{compN2Ha}
\beq
G^2_{-}  = \psi_-^\mu \d_=X^\nu J_{-\mu\nu} -
\frac{i}{3} \psi_-^\mu \psi^\nu_- \psi^\rho_- J^{\lambda}_{-\mu}
J^\sigma_{-\nu} J^\gamma_{-\rho}
H_{\lambda\sigma\gamma} ,
\eeq{compN2Hb}
\beq
J_+ = \psi_+^\mu \psi_+^\nu J_{+\mu\nu},
\eeq{R1cur}
\beq
J_{-} = \psi_-^\mu \psi_-^\nu J_{-\mu\nu} .
\eeq{compN2Hcrsym}
If we want to preserve the maximal possible amount of the bulk
supersymmetries in the presence of a boundary, then in addition to
the conditions
(\ref{bulkcurbc}) we need to simultaneously require the following
boundary conditions for the currents $G^2_{\pm}$ and $J_{\pm}$,
\beq
[G^2_+ - \eta_2 G^2_-]_{\sigma =0,\pi} =0 ,
\,\,\,\,\,\,\,\,\,\,\,\,\,\,\,\,\,\,\,
[J_+ - (\eta_1 \eta_2) J_- ]_{\sigma = 0,\pi} = 0 ,
\eeq{G2cond}
where $\eta_2 = \pm 1$.
As a result, the conditions
(\ref{RgRcond1}) and (\ref{ALZ320}) on $R^\mu_{\,\,\nu}$
must be supplemented with the following,
\beq
J^\mu_{-\lambda} R^\lambda_{\,\,\nu} = (\eta_1 \eta_2 )
R^\mu_{\,\,\lambda} J^\lambda_{+\nu} ,
\eeq{JRJRJR}
\beq
J^\lambda_{+[\mu} H_{|\lambda|\nu|\sigma|}
P^\sigma_{\,\,\rho]} + J^\lambda_{+[\mu} R^\sigma_{\,\,|\lambda|}
P^\gamma_{\,\,\nu} R^\delta_{\,\,\rho]}
H_{\sigma\gamma\delta} - J^\lambda_{+[\mu} R^{\phi}_{\,\,|\lambda}
\eta_{\phi\gamma} \nabla_{\sigma|} R^\gamma_{\,\,\rho}
P^\sigma_{\,\,\nu]}= 0 ,
\eeq{extracond}
where the case $(\eta_1 \eta_2 )= 1$ corresponds to B-type and $(\eta_1
\eta_2 )=-1$ to A-type models.

It was shown in \cite{LZ1} that (\ref{extracond}) is automatically
satisfied, given the other three conditions, (\ref{RgRcond1}),
(\ref{ALZ320}) and (\ref{JRJRJR}). Nevertheless, it is interesting
to note that (\ref{extracond})
corresponds to the invariance of the two-fermion term in
(\ref{boscond}) under the appropriate combination of
$U(1)$-symmetries.

Geometrically the $N=2$ superconformal boundary conditions should be
interpreted in terms of submanifolds of special types. However, the
problem has not been solved in all generality (some observations on
the subject are made in \cite{LZ1}). Only $N=2$ D-branes on K\"ahler
manifolds (i.e., $J^{\mu}_{+\nu} = \pm J^{\mu}_{-\nu}$) are well
understood in terms of symplectic geometry. Unfortunately this case is
irrelevant to the WZW models since group manifolds are never
K\"ahler.

\Section{The ${\mathbf N=1}$ WZW model}
\label{N1WZW}

We now analyse the $N=1$ boundary conditions given in
section~\ref{N1review} for the WZW model. We begin by reviewing some
basics pertaining to WZW models, in the process introducing some
notation that will be useful in the analysis.

\subsection{Preliminaries}

The WZW models represent a special class of non-linear sigma models
defined over a group manifold ${\cal M}$ of some Lie group ${\cal G}$.
The isometry group ${\cal G}\times {\cal G}$ is generated by the left-
and right-invariant Killing vectors $l^\mu_A$ and $r^\mu_A$
respectively, where $A=1,2,...,\dim\,{\cal G}$. They satisfy
\beq
\{ l_A, l_B \} = f_{AB}^{\,\,\,\,\,\,\,\,C} l_C,\,\,\,\,\,\,\,\,\,\,\,\,
\{ r_A, r_B \} = -f_{AB}^{\,\,\,\,\,\,\,\,C} r_C,\,\,\,\,\,\,\,\,\,\,\,\,
\{ l_A, r_B\}=0 ,
\eeq{liebrakalg}
where $\{\, \cdot \, ,\, \cdot \, \}$
is the Lie bracket for vector fields.  We restrict
ourselves to semi-simple Lie groups, so that the Cartan-Killing metric
$\eta_{AB}$ has an inverse $\eta^{AB}$ and can be used to raise and
lower Lie algebra indices. Both $l^\mu_A$ and $r^\mu_A$ can be
regarded as vielbeins, with inverses $l^A_\mu$ and $r^A_\mu$,
respectively. To define the sigma model, we choose the invariant
metric
\beq
g_{\mu\nu} = \frac{1}{\rho^2}\, l^A_\mu l^B_\nu \eta_{AB} =
\frac{1}{\rho^2}\, r^A_\mu r^B_\nu \eta_{AB} ,
\eeq{metrinv}
while $H_{\mu\nu\rho}$ is proportional to the structure constants of
the corresponding Lie algebra ${\bf g}$,
\beq
H_{\mu\nu\rho} = \frac{1}{2} k\, l^A_\mu l^B_\nu l^C_\rho f_{ABC} =
\frac{1}{2} k\, r^A_\mu r^B_\nu r^C_\rho f_{ABC}.
\eeq{deftors}
Here $\rho$ and $k$ are constants, and $k$ must satisfy a quantisation
condition.  If $\rho^2 = \pm 1/k$, then $H_{\mu\nu\rho}$ is the
parallelising torsion on the group manifold and this is also precisely
the relation between the coupling constants that holds at the
conformal fixed point of the beta-functions. Since we are interested
in the conformal model, we set $\rho^2=1/k$ in the following
discussion. Moreover, since $k$ appears only as an overall factor in
our calculations, we may set $k=1$.

We thus study the sigma model
(\ref{sigmamodel}) with $g_{\mu\nu}$ and $H_{\mu\nu\rho}$ given by
(\ref{metrinv}) and (\ref{deftors}). From the above properties follows
that the left- and right-invariant Killing vectors satisfy
\beq
\nabla^{(-)}_\rho l^\mu_A =0,\,\,\,\,\,\,\,\,\,\,\,\,\,\,
\,\,\,\,\nabla^{(+)}_\rho r^\mu_A=0,
\eeq{covconKV}
where $\nabla_\rho^{(\pm)}$ are the affine connections defined in
(\ref{definGammapm}).  The relations (\ref{covconKV}) are
the Cartan-Maurer equations for our group manifold.
They imply the existence of chiral
(antichiral) Lie algebra valued currents,
\beq
{\cal J}^A_- = l^A_\mu D_- \Phi^\mu,\,\,\,\,\,\,\,\,\,\,\,\,\,\,
{\cal J}^A_+ = - r^A_\mu D_+ \Phi^\mu,
\eeq{fefgroupcur}
obeying $D_{\mp} {\cal J}^A_{\pm} =0$. The components of these
currents are defined as
\beq
j_{\pm}^A = {\cal J}^A_{\pm}|,\,\,\,\,\,\,\,\,\,\,\,\,\,\,\,\,
k^A_{\pp} = -iD_{\pm} {\cal J}^A_{\pm}| .
\eeq{compgrcur}
It is important to notice that there are two different sets of
bosonic affine currents, ${\cal K}^A_{\pp}$ and $k^A_{\pp}$,
related to each other by
\beq
\begin{array}{l}
  k_=^A= {\cal K}^A_= + \frac{i}{2}
  f_{BC}^{\,\,\,\,\,\,\,\,A} j_-^B j_-^C , \\ \\
  k_\+^A= {\cal K}^A_\+ + \frac{i}{2}
  f_{BC}^{\,\,\,\,\,\,\,\,A} j_+^B j_+^C ,
\end{array}
\eeq{kcurK}
where 
\beq
{\cal K}_=^A \equiv l^A_\mu \d_= X^\mu,\,\,\,\,\,\,\,\,\,\,\,\,
{\cal K}_\+^A \equiv - r^A_\mu \d_\+ X^\mu  .
\eeq{defcalK}
In terms of group elements $g$, they correspond to the currents
${\cal K}_= = g^{-1} dg$ and ${\cal K}_\+ = dg\, g^{-1}$.
However, here we will stick to the coordinate representation.

Note that, on a group manifold with metric (\ref{metrinv}) and torsion
(\ref{deftors}), with $\rho^2 = 1/k$,
the four-fermion term in the action (\ref{sigmamodel})
vanishes. Using the Lie algebra valued
fermion fields defined in (\ref{compgrcur}),
\beq
j_-^A = l^A_\mu \psi^\mu_-,\,\,\,\,\,\,\,\,\,\,\,\,\,\,\,\,
j_+^A = - r^A_\mu \psi^\mu_+,
\eeq{defferm234}
we can rewrite (\ref{sigmamodel}) as
\beq
S=  \int d^2\xi\,\,\left [ \d_\+ X^\mu \d_= X^\nu E_{\mu\nu} +
  i\, j_-^A \, \eta_{AB} \, \d_{\+} j_-^B   + i\, j_+^A \, \eta_{AB}\, \d_{=}
  j_+^B \right ] ,
\eeq{newactjj}
where the bosonic part can be written in terms of the group
elements as usual.

\subsection{Boundary conditions}

Here we discuss the $N=1$ superconformal boundary conditions which we
reviewed in section~\ref{N1review}, for the WZW models. The conditions
(\ref{fermans}) and (\ref{boscond}) can be rewritten in terms of the
affine bosonic and fermionic currents, as
\beq
j_-^A = \eta_1 R^A_{\,\,\,B} j_+^B ,
\eeq{fermibcaffc}
\beq
{\cal K}_=^A - R^A_{\,\,\,B} {\cal K}_\+^B 
+ \frac {i}{2} \left (  f_{SM}^{\,\,\,\,\,\,\,\,A} R^S_{\,\,D}
  R^M_{\,\,L} -  f_{DL}^{\,\,\,\,\,\,\,\,C} R^A_{\,\,\,C}
  - 2 {\cal L}_D R^A_{\,\,\,L} \right )
j_+^D j_+^L = 0 .
\eeq{bosbcafc}
Here we have defined
\beq
R^\mu_{\,\,\,\nu} \equiv - l_A^\mu R^A_{\,\,\,B} r^B_\nu,
\,\,\,\,\,\,\,\,\,\,\,\,\,\,\,
{\cal L}_D R^A_{\,\,\,L} \equiv k^\mu_D \d_\mu R^A_{\,\,\,L} ,
\,\,\,\,\,\,\,\,\,\,\,\,\,\,\,
k_A^\mu \equiv r^\mu_A - R^L_{\,\,\,A} l^\mu_L .
\eeq{Liedefs}
Thus $R^A_{\,\,\,B}:
{\cal G} \rightarrow {\mathbf g} \otimes {\mathbf g}^*$ is a map
from the group to the tensor product of the Lie algebra with its dual.

In terms of Lie algebra quantities, the
conditions (\ref{RgRcond1}) and (\ref{ALZ320}) can
similarly be rewritten as
\beq
R^C_{\,\,\,A} \eta_{CD} R^D_{\,\,\,B} = \eta_{AB} ,
\eeq{presmetrWZW}
\beq
f_{ABC} - R^D_{\,\,\,A}
R^L_{\,\,\,B} R^M_{\,\,\,C} f_{DLM} = - \eta_{LM} R^M_{\,\,\,[C}
{\cal L}_{A} R^L_{\,\,\,B]}  .
\eeq{N1suprWZW}
The condition (\ref{N1suprWZW}) is a first-order differential
equation for the gluing matrix $R^A_{\,\,\,B}$.
The most obvious solution of eq.\ (\ref{N1suprWZW}) is a constant
matrix $R^A_{\,\,\,B}$, i.e., ${\cal L}_C R^A_{\,\,\,B}= 0$.  In this
case $R^A_{\,\,\,B}$ corresponds to a Lie algebra automorphism of
${\mathbf g}$. The boundary conditions (\ref{fermibcaffc}) and
(\ref{bosbcafc}) are then local in terms of the affine currents. This
case is best known and geometrically corresponds to (twisted)
conjugacy classes \cite{Felder,Stanciu:1999id}.  Note that
(\ref{N1suprWZW}) allows also nonconstant $R^A_{\,\,\,B}$ to be Lie
algebra automorphisms, as long as they satisfy ${\cal L}_{[A}
R^C_{\,\,\,B]} =0$.

Although it is plausible from a technical point of view to write the
boundary conditions in a local form for the affine currents, there is
no physical reason why all (in some sense) reasonable conditions
should be local in these currents.  At the level of sigma models,
however, locality in terms of $X^\mu$ and $\psi^\mu_{\pm}$ is
necessary for a geometrical interpretation.

Another interesting point is that, if we require the two-fermion term
in (\ref{bosbcafc}) to be absent, then combining this with
(\ref{N1suprWZW}), we find that $R^A_{\,\,\,B}$ is constant along
$\pi$, that is, $\pi^\mu_{\,\,\,\nu} \d_\mu R^A_{\,\,\,B} =0$.  Thus
it corresponds to a Lie algebra automorphism along $\pi$.

\Section{The ${\mathbf N=2}$ WZW model}
\label{N2WZW}

\subsection{Preliminaries}

The problem of $N=2$ supersymmetry for WZW models was first addressed
in \cite{Spindel:1988nh,Spindel:1988sr,Sevrin:1988ps}.  However, we
will not follow the original presentation; instead, the basic approach
used here, as well as some of the results, may be found in \cite{LZ2}.

The complex structures $J^\mu_{\pm\nu}$ are, respectively, left- and
right-invariant, and are of the form
\beq
J^\mu_{-\nu} = l^\mu_A J^A_{\,\,B} l^B_\nu,\,\,\,\,\,\,\,\,\,\,\,\,\,\,\,
J^\mu_{+\nu} = r^\mu_A \tilde{J}^A_{\,\,B} r^B_\nu ,
\eeq{supersymemN2c}
where $J^A_{\,\,B}$ and $\tilde{J}^A_{\,\,B}$ are constant
matrices acting on the Lie algebra. Restricting attention to the
left-invariant complex structure $J^A_{\,\,B}$, bulk supersymmetry
imposes the following constraints,
\beq
J^A_{\,\,C} J^C_{\,\,B}= -\delta^A_{\,\,B} ,
\eeq{Jgrppro}
\beq
J^C_{\,\,A} \eta_{CD} J^D_{\,\,B} = \eta_{AB} ,
\eeq{hemrJAB}
\beq
f_{ABC} = J^D_{\,\,A} J^L_{\,\,B} f_{DLC} + J^D_{\,\,B} J^L_{\,\,C} f_{DLA}
+ J^D_{\,\,C} J^L_{\,\,A} f_{DLB} .
\eeq{HJHHAB}
Thus we need to construct a $J^A_{\,\,B}$ on the Lie algebra ${\bf g}$
satisfying (\ref{Jgrppro})--(\ref{HJHHAB}). This is possible only for
even-dimensional Lie algebras. $J^A_{\,\,B}$ has eigenvalues $\pm i$,
and we choose a basis $T_A=(T_a,T_{\bar{a}})$ on the Lie algebra ${\bf
  g}$ such that $J^A_{\,\,B}$ is diagonal: $J^a_{\,\,b}= i
\delta^a_{\,\,b}$, $J^{\bar{a}}_{\,\,\bar{b}}= -i
\delta^{\bar{a}}_{\,\,\bar{b}}$.  In this basis eq.\ (\ref{hemrJAB})
leads to $\eta_{ab}=\eta_{\bar{a}\bar{b}}=0$, and (\ref{HJHHAB}) gives
$f_{abc}=f_{\bar{a}\bar{b}\bar{c}}=0$.  Taken together, this implies
that $f_{ab}^{\,\,\,\,\,\,{\bar{c}}}=0$ and
$f_{\bar{a}\bar{b}}^{\,\,\,\,\,\,c}=0$, so that the two sets of
generators $\{T_a\}$ and $\{ T_{\bar{a}}\}$ form Lie
subalgebras of ${\bf g}$, call them
${\bf g_+}$ and ${\bf g_-}$, respectively. These
subalgebras are maximally isotropic subspaces with respect to
$\eta_{AB}$. Thus the complex structures on the even-dimensional group
are related to a decomposition of the Lie algebra ${\bf g}$ into two
maximally isotropic subalgebras with respect to $\eta_{AB}$, such that
${\bf g} = {\bf g_-} \oplus {\bf g_+}$ as a vector space.  Such a
structure is called a Manin triple $({\bf g}, {\bf g_-}, {\bf g_+})$,
and was initially introduced by Drinfeld in the context of completely
integrable systems and quantum groups \cite{Drinfeld:in}.  The
relevance of Manin triples to $N=2$ supersymmetry on group manifolds
was pointed out in \cite{Parkhomenko:dq}.

In the general situation the $N=(2,2)$ WZW model would be equipped
with a left Manin triple $({\bf g}, {\bf g_-}, {\bf g_+})$, and a
right Manin triple $({\bf g}, {\bf \tilde{g}_-}, {\bf \tilde{g}_+})$.
However, both left and right Manin triples are defined with respect to
the same ad-invariant bilinear nondegenerate form $\eta_{AB}$.  If the
left and right Manin triples are the same (i.e.,
$J^A_{\,\,B}=\tilde{J}^A_{\,\,B}$ modulo a Lie algebra automorphism),
then the bivector $(J_+ - J_-)g^{-1}$ defines a Poisson-Lie structure.
Here we consider the situation where the left and right Manin triples
are the same; for different left and right Manin triples the
generalisation is straightforward.

\subsection{Boundary conditions}

In this section we study the implications of the $N=2$ boundary
conditions (reviewed in section~\ref{N2review}) for the WZW model.
The currents may be written in
terms of Lie algebra quantities (see appendix~\ref{a:N2currents}),
and the boundary conditions corresponding to (\ref{JRJRJR}) and
(\ref{extracond}) read
\beq
R^B_{\,\,A} J^C_{\,\,B} = (\eta_1\eta_2) J^B_{\,\,A} R^C_{\,\,B} ,
\eeq{ABtypRJ}
\beq
J_{NM} R^N_{\,\,[A} {\cal L}_{B} R^M_{\,\,C]} = 0 .
\eeq{newconN2jrj}
We first analyse the B-type conditions, $(\eta_1 \eta_2 )= 1$.
In this case eq.\ (\ref{ABtypRJ}) implies that,
in our chosen basis, $R^a_{\,\,\bar{b}} = 
R^{\bar{a}}_{\,\,b} = 0$.
Then we find from the $N=1$ condition (\ref{N1suprWZW}) that
(recall that $f_{abc}=f_{\bar{a}\bar{b}\bar{c}}=0$)
\ber
& & f_{ab\bar{c}} - f_{dl\bar{m}}
R^d_{\,\,a} R^l_{\,\,b} R^{\bar{m}}_{\,\,\bar{c}} =
-  \eta_{l\bar{m}} R^{\bar{m}}_{\,\,\bar{c}}
\left( {\cal L}_{[a} R^l_{\,\,b]} \right) ,   \label{BtypeN1} \\
& & f_{\bar{a}\bar{b}c} - f_{\bar{d}\bar{l}m}
R^{\bar{d}}_{\,\,\bar{a}} R^{\bar{l}}_{\,\,\bar{b}}
R^{m}_{\,\,c} =
-  \eta_{\bar{l}m} R^{m}_{\,\,c} \left( {\cal L}_{[\bar{a}}
  R^{\bar{l}}_{\,\,\bar{b}]} \right) .
\eer{solBtyRetaR}
Using (\ref{presmetrWZW}), we can write this as
\ber
&& f_{ab}^{\,\,\,\,\,\,s} R^l_{\,\,s} -
f_{mn}^{\,\,\,\,\,\,l}  R^m_{\,\,a}  R^n_{\,\,b}
= - {\cal L}_{[a} R^l_{\,\,b]} , \label{holRLReq}\\
&& f_{\bar{a}\bar{b}}^{\,\,\,\,\,\,\bar{s}} R^{\bar{l}}_{\,\,\bar{s}} -
f_{\bar{m}\bar{n}}^{\,\,\,\,\,\,\bar{l}}
R^{\bar{m}}_{\,\,\bar{a}}  R^{\bar{n}}_{\,\,\bar{b}}
= - {\cal L}_{[\bar{a}} R^{\bar{l}}_{\,\,\bar{b}]} .
\eer{newfRLR}

For the A-type conditions, $(\eta_1 \eta_2 )=-1$,
eq.\ (\ref{ABtypRJ}) implies that $R^a_{\,\,b} =
R^{\bar{a}}_{\,\,\bar{b}} =0$.
The relations corresponding to (\ref{BtypeN1})
and (\ref{solBtyRetaR}) for A-type are then
\ber
& &
f_{ab\bar{c}}-f_{\bar{d}\bar{l}m}R^{\bar{d}}_aR^{\bar{l}}_bR^m_{\bar{c}}
=- \eta_{l\bar{m}}R^l_{\bar{c}} \left( {\cal L}_{[a}R^{\bar{m}}_{b]} \right),
\label{AtypeN1} \\
& & f_{\bar{a}\bar{b}c} - f_{dl\bar{m}}
R^{d}_{\,\,\bar{a}} R^{l}_{\,\,\bar{b}}
R^{\bar{m}}_{\,\,c} =
-  \eta_{l\bar{m}} R^{\bar{m}}_{\,\,c} \left( {\cal L}_{[\bar{a}}
  R^{l}_{\,\,\bar{b}]} \right) ,
\eer{solAtyRetaR}
or equivalently,
\ber
&& f_{ab}^{\,\,\,\,\,\,s} R^{\bar{l}}_{\,\,s} -
f_{\bar{m}\bar{n}}^{\,\,\,\,\,\,\bar{l}}
R^{\bar{m}}_{\,\,a}  R^{\bar{n}}_{\,\,b}
= - {\cal L}_{[a} R^{\bar{l}}_{\,\,b]} , \label{holRLReqA}\\
&& f_{\bar{a}\bar{b}}^{\,\,\,\,\,\,\bar{s}} R^{l}_{\,\,\bar{s}} -
f_{mn}^{\,\,\,\,\,\,l}
R^{m}_{\,\,\bar{a}}  R^{n}_{\,\,\bar{b}}
= - {\cal L}_{[\bar{a}} R^{l}_{\,\,\bar{b}]} .
\eer{newfRLRA}

Let us analyse the B-type equation (\ref{holRLReq}) in detail (the
analogous consideration can be applied to (\ref{newfRLR}), as well as
to the A-type equations). To understand the geometrical content of this
condition, we first consider the vector $k_a^\mu = r^\mu_a -
R^l_{\,\,a} l^\mu_l$, with respect to which the Lie derivative in
(\ref{holRLReq}) is taken.
We can ask when $k_a^\mu$ satisfies the Lie algebra bracket
\beq
\{ k_a, k_b \} = -f_{ab}^{\,\,\,\,\,\,\,\,c} k_c .
\eeq{algkvect}
To obtain the answer, we use that the vectors
$l^\mu_a$ and $r^\mu_a$ generate the algebra ${\mathbf g_-}$,
\beq
\{ l_a, l_b \} = f_{ab}^{\,\,\,\,\,\,\,\,c}
l_c,\,\,\,\,\,\,\,\,\,\,\,\, \{ r_a, r_b \} =
-f_{ab}^{\,\,\,\,\,\,\,\,c} r_c,\,\,\,\,\,\,\,\,\,\,\,\, \{ l_a,
r_b\}=0 .
\eeq{algegplus}
We find that $k_a^\mu$ satisfies (\ref{algkvect})
if and only if the following holds,
\beq
f_{ab}^{\,\,\,\,\,\,s} R^l_{\,\,s} -
f_{mn}^{\,\,\,\,\,\,l}  R^m_{\,\,a}  R^n_{\,\,b} 
= - {\cal L}_{[a} R^l_{\,\,b]} .
\eeq{eqautgener}
This is precisely the condition (\ref{holRLReq}).

If $R^l_{\,\,b}$ is constant then we find that it
corresponds to an automorphism of the Lie algebra ${\mathbf g_-}$.
In this case $k_a^\mu$ has a clear geometrical interpretation: it
generates a twisted adjoint action of the group ${\cal G}$
\cite{Stanciu:1999id}.

However, we are interested in the general case of nonconstant
$R^l_{\,\,b}$. It is clear from the analysis above that this
generalisation occurs in a controlled way. It simply modifies the
vector $k_a^\mu$ by allowing $X$-dependence of the map $R^l_{\,\,b}$
in the definition of $k_a^\mu$.  Moreover, the deviation of
$R^l_{\,\,b}$ from being an automorphism is governed by
(\ref{eqautgener}).

In conclusion, we see that the most general $N=2$ superconformal
boundary conditions are a very precise generalisation of the
well-known ones that correspond to constant Lie algebra automorphisms.
In geometrical terms, the resulting D-branes are related in a specific
way to the known D-branes that correspond to conjugacy classes of
${\cal G}$.

It is interesting to examine the integrability conditions for
eqs.\ (\ref{holRLReq}),
(\ref{newfRLR}), (\ref{holRLReqA}) and (\ref{newfRLRA}).
We focus on (\ref{holRLReq}), hit it on both sides with
${\cal L}_c$ and antisymmetrise in $a$, $b$ and $c$, to get
\beq
f_{[ab}^{\,\,\,\,\,\,d} {\cal L}_{c]} R^l_{\,\,d} -
f_{mn}^{\,\,\,\,\,\,l}  {\cal L}_{[c} \left( R^m_{\,\,a}  R^n_{\,\,b]}
\right)  = - 2 {\cal L}_{[c} {\cal L}_{a} R^l_{\,\,b]} .
\eeq{Lhit}
Using that ${\cal L}_{[a}
{\cal L}_{b]} = - f_{ab}^{\,\,\,\,\,\,d} {\cal L}_d$
(which follows from (\ref{algkvect})), we then rewrite the
right-hand side of (\ref{Lhit}) as
\beq
f_{[ab}^{\,\,\,\,\,\,d}{\cal L}_{|d|}R^l_{\,\,c]} ,
\eeq{RHS1}
which, using (\ref{holRLReq}) as well as the Jacobi identity,
expands to
\beq
f_{[ab}^{\,\,\,\,\,\,d}{\cal L}_{c]}R^l_{\,\,d}
+ f_{mn}^{\,\,\,\,\,\,l} f_{[ab}^{\,\,\,\,\,\,d}
R^m_{\,\,|d|}R^n_{\,\,c]} .
\eeq{RHS2}
On the other hand expanding the
left-hand side of (\ref{Lhit}), and again using the Jacobi identity, yields
\beq
f_{[ab}^{\,\,\,\,\,\,d}{\cal L}_{c]}R^l_{\,\,d}
+ 2 f_{mn}^{\,\,\,\,\,\,l} f_{[ab}^{\,\,\,\,\,\,d}
R^m_{\,\,|d|}R^n_{\,\,c]}.
\eeq{LHS}
Thus (\ref{Lhit}) is satisfied if and only if
(\ref{RHS2}) and (\ref{LHS}) are equal, i.e.\ the condition for
(\ref{holRLReq}) to be integrable is
\beq
f_{mn}^{\,\,\,\,\,\,l} f_{[ab}^{\,\,\,\,\,\,d}
R^m_{\,\,|d|}R^n_{\,\,c]} =0.
\eeq{WZWinteg}
This equation arises in Lie algebra cohomology, and is solved by
\beq
R^m_{\,\,s} f_{mn}^{\,\,\,\,\,\,l} R^n_{\,\,p} = f_{sp}^{\,\,\,\,\,\,q}
T^l_{\,\,q},
\eeq{WZWsolution}
for some $T^l_{\,\,q}$.  For simple Lie algebras the corresponding
cohomology is trivial and this is the {\em only} solution to
(\ref{WZWinteg}).  If $T^l_{\,\,q} = R^l_{\,\,q}$, then
(\ref{WZWsolution}) implies that $R^l_{\,\,q}$ is a Lie algebra
automorphism, and otherwise the difference $f_{ab}^{~~q}(T^l_{\,\,q}
-R^l_{\,\,q})={\cal L}_{[a}R^l_{\,\,b]}$ is a measure of the extent
to which it fails to be such an automorphism.

With the appropriate change of indices these results also apply to the
remaining equations (\ref{newfRLR}), (\ref{holRLReqA}) and
(\ref{newfRLRA}).

\Section{Discussion}
\label{discussion}

In this letter we have analysed the local classical superconformal
boundary conditions for non-linear sigma models with specific target
spaces, namely group manifolds.  Information about the local boundary
conditions is encoded in the gluing matrix $R^A_{\,\,B} :{\cal G}
\rightarrow {\mathbf g} \otimes {\mathbf g}^*$.  Using the special
features of WZW models we found the conditions that $R^A_{\,\,B}$
has to obey to preserve superconformal symmetry on the boundary.
These conditions are a set of first-order differential equations for
the gluing matrix. In the case of $N=2$ supersymmetry, these equations
become especially interesting, and we also analysed their
integrability. The geometrical significance of these results remains
to be determined.

Our analysis is entirely classical and the well-known D-branes
corresponding to Lie algebra automorphims are special solutions of our
conditions.

At the quantum level the WZW models are not rational with respect to
(super)conformal symmetry and therefore it is problematic to analyse
the (super)conformal boundary conditions by means of a rational CFT.
Thus, at present, at the quantum level, a full description of D-branes
on group manifolds is still an open problem. It would be interesting
to see if there are classical branes with nonconstant $R^A_{\,\,B}$
which have a quantum counterpart.

\bigskip

\bigskip

{\bf Acknowledgments}: CA is grateful to Queen Mary, University of
London, for hospitality while part of this work was carried out, and
acknowledges support in part by EU contract HPMT-CT-2000-00025.  UL
acknowledges support in part by EU contract HPNR-CT-2000-0122 and by
VR grant 650-1998368. MZ is grateful to Anton Alekseev, Andrea
Cappelli and Pierre Vanhove for useful discussions on this and related
subjects, and acknowledges support in part by EU contract
HPRN-CT-2002-00325.

\appendix

\section{N=2 currents for the WZW model}
\label{a:N2currents}

Here we list the components of the $N=2$ stress tensor $T_{\pm\pm}$,
supersymmetry currents $G^1_{\pm}$ and $G^2_{\pm}$, and R-symmetry
currents $J_{\pm}$, in terms of the Lie algebra valued affine currents
$j^A_{\pm}$ and ${\cal K}_{\pp}^A$.  These are obtained by
substituting the relations (\ref{metrinv}), (\ref{deftors}),
(\ref{defcalK}) and (\ref{defferm234}) into the expressions
(\ref{Gp})--(\ref{Tm}) and (\ref{compN2Ha})--(\ref{compN2Hcrsym}).
\beq
T_{++} = {\cal K}_{\+}^A \eta_{AB} {\cal K}_{\+}^B + i j_+^A \eta_{AB} \d_\+
j^B_+ ,
\eeq{comp3}
\beq
T_{--} =  {\cal K}_{=}^A \eta_{AB} {\cal K}_{=}^B + i j_-^A \eta_{AB} \d_=
j^B_- ,
\eeq{comp4N2H}
\beq
G^1_+ =  j_+^A \eta_{AB} {\cal K}_{\+}^B + \frac{i}{6} j_+^A j_+^B j_+^C f_{ABC}
,
\eeq{comp1N2H}
\beq
G^1_- = j_-^A \eta_{AB} {\cal K}_{=}^B + \frac{i}{6} j_-^A j_-^B j_-^C f_{ABC}
,
\eeq{comp2N2H}
\beq
G^2_+ =  j_+^A J_{AB} {\cal K}_{\+}^B - \frac{i}{6} j_+^A j_+^B j_+^C
J^D_{\,\,A}
J^L_{\,\,B} J^M_{\,\,C} f_{DLM} ,
\eeq{AcompN2Ha}
\beq
G^2_- = j_-^A J_{AB} {\cal K}_{=}^B - \frac{i}{6} j_-^A j_-^B j_-^C
J^D_{\,\,A}
J^L_{\,\,B} J^M_{\,\,C} f_{DLM} ,
\eeq{AcompN2Hb}
\beq
J_+ = j_+^A j_+^B J_{AB},\,\,\,\,\,\,\,\,\,\,\,\,\,\,\,\,
J_- = j_-^A j_-^B J_{AB} .
\eeq{AcompN2Hcrsym}

\end{document}